\documentclass[12pt]{article}

\usepackage{amsmath,amscd,amsthm,amssymb}
\usepackage{fancyhdr,graphics,pifont}
\usepackage{floatflt,subfigure,epsfig,moreverb,multicol,lscape}
\usepackage{supertabular,tabularx,multirow,bar}
\usepackage{psfrag,afterpage,curves,epic, bbm}
\usepackage[english]{babel}

\newcommand{\supp}{\operatorname{supp}}

\newcommand{\matr}[1]{\mathbf{#1}}
\newcommand{\vect}[1]{\mathbf{#1}}
\newcommand{\code}[1]{\mathcal{#1}}
\newcommand{\set}[1]{\mathcal{#1}}

\newcommand{\R}{\mathbb{R}}
\newcommand{\Rp}{\mathbb{R}_{+}}
\newcommand{\Rpp}{\mathbb{R}_{++}}

\newcommand{\PG}[2]{\operatorname{PG}(#1,#2)}

\newcommand{\PGq}{\operatorname{PG}(2,q)}
\newcommand{\codePGq}{\code{C}_{\PGq}}

\newcommand{\wcol}{w_{\mathrm{col}}}
\newcommand{\wrow}{w_{\mathrm{row}}}
\newcommand{\defeq}{\triangleq}

\newcommand{\vlambda}{\boldsymbol{\lambda}}
\newcommand{\vomega}{\boldsymbol{\omega}}

\newcommand{\sonenorm}[1]{\lVert #1 \rVert_1}
\newcommand{\onenorm}[1]{\lVert #1 \rVert_1}
\newcommand{\twonorm}[1]{\lVert #1 \rVert_2}

\newcommand{\vx}{\vect{x}}

\theoremstyle{definition}
\newtheorem{Definition}{Definition}
\newtheorem{Example}[Definition]{Example}

\theoremstyle{plain}
\newtheorem{Lemma}[Definition]{Lemma}
\newtheorem{Theorem}[Definition]{Theorem}
\newtheorem{Corollary}[Definition]{Corollary}

\newtheorem{Conjecture}[Definition]{Conjecture}

\theoremstyle{definition}

\newenvironment{Proof}%
  {\noindent \emph{Proof:}}{\hfill$\square$}

\newcommand{\edefinition}{\hfill$\square$}

\newcommand{\eexample}{\hfill$\square$}

\newcommand{\fc}[1]{\set{#1}}
\newcommand{\fch}[2]{\set{#1}(\matr{#2})}
\newcommand{\Mps}{\set{M}_{\mathrm{p}}}

\newcommand{\wps}{w_{\mathrm{p}}}
\newcommand{\wpsAWGNC}{w_{\mathrm{p}}^{\mathrm{AWGNC}}}
\newcommand{\wpsBSC}{w_{\mathrm{p}}^{\mathrm{BSC}}}
\newcommand{\wpsBEC}{w_{\mathrm{p}}^{\mathrm{BEC}}}
\newcommand{\wpsmin}{w_{\mathrm{p}}^{\mathrm{min}}}

\newcommand{\wpsBSCmin}{w_{\mathrm{p}}^{\mathrm{BSC,min}}}
\newcommand{\wpsBECmin}{w_{\mathrm{p}}^{\mathrm{BEC,min}}}

\newcommand{\type}{t}
\newcommand{\vtype}{\vect{t}}

\begin{document}

\renewcommand{\textfraction}{0}

\title{On Minimal Pseudo-Codewords of \\ Tanner Graphs from Projective
       Planes\footnote{The first author was supported by NSF Grant ATM
       02-96033, by DOE SciDAC, and by ONR Grant N00014-00-1-0966.  The second
       author was supported by NSF Grant ITR 02-05310. This paper is a
       slightly reformulated version of the paper that appeared in the
       proceedings of the 43rd Allerton Conference on Communications, Control,
       and Computing, Allerton House, Monticello, Illinois, USA, Sept.~28--30,
       2005.}}

\author{\normalsize
  \normalsize Pascal O.~Vontobel \\ 
  \small      Dept.~of ECE \\
  \small      University of Wisconsin \\
  \small      Madison, WI 53706, USA \\
  \small      \texttt{vontobel@ece.wisc.edu}
  \and
  \normalsize Roxana Smarandache \\
  \small      Dept.~of Mathematics \\
  \small      University of Notre Dame \\
  \small      Notre Dame, IN 46556, USA \\
  \small      \texttt{rsmarand@nd.edu}
}

\date{}

\maketitle

\begin{abstract}
  We would like to better understand the fundamental cone of Tanner graphs
  derived from finite projective planes. Towards this goal, we discuss bounds
  on the AWGNC and BSC pseudo-weight of minimal pseudo-codewords of such
  Tanner graphs, on one hand, and study the structure of minimal
  pseudo-codewords, on the other.
\end{abstract}

\section{Introduction}

In this paper we focus solely on certain families of codes based on finite
projective planes. More precisely, the codes under investigation are the
families of codes that were called type-I PG-LDPC codes
in~\cite{Lucas:Fossorier:Kou:Lin:00:1, Kou:Lin:Fossorier:01:1}, see
also~\cite{Vontobel:Smarandache:Kiyavash:Teutsch:Vukobratovic:05:1}. They are
defined as follows. Let $q \defeq 2^s$ for some positive integer $s$ and
consider a (finite) projective plane $\PGq$ (see e.g.~\cite{Hirschfeld:79:1,
Batten:97}) with $q^2 + q + 1$ points and $q^2 + q + 1$ lines: each point lies
on $q+1$ lines and each line contains $q+1$ points.\footnote{Note that the
``$2$'' in $\PGq$ stands for the dimensionality of the geometry, which in the
case of planes is $2$.} A standard way of associating a parity-check matrix
$\matr{H}$ of a binary linear code to a finite geometry is to let the set of
points correspond to the columns of $\matr{H}$, to let the set of lines
correspond to the rows of $\matr{H}$, and finally to define the entries of
$\matr{H}$ according to the incidence structure of the finite geometry. In
this way, we can associate to the projective plane $\PGq$ the code $\codePGq$
with parity-check matrix $\matr{H} \defeq \matr{H}_{\PGq}$. It turns out
that this code has block length $n = q^2 + q + 1$, dimension $n - 3^s - 1$,
and minimum Hamming distance $q + 2$. The parity-check matrix
$\matr{H}_{\PGq}$ has size $n \times n$ and it has uniform column weight
$\wcol = q + 1$ and uniform row weight $\wrow = q + 1$. Moreover, this code
has the nice property that with an appropriate ordering of the columns and
rows, the parity-check matrix is a circulant matrix, meaning that $\codePGq$
is a cyclic code. This fact can e.g.~be used for efficient encoding. Such
symmetries can also substantially simplify the analysis. Note that the
automorphism group of $\codePGq$ contains many more automorphisms besides the
cyclic-shift-automorphism implied by the cyclicity of the code.

In this paper we continue the investigations started
in~\cite{Vontobel:Smarandache:Kiyavash:Teutsch:Vukobratovic:05:1} related to
these codes. Our goal is to improve our knowledge about the fundamental
cone~\cite{Koetter:Vontobel:03:1, Feldman:Wainwright:Karger:05:1} of the
parity-check matrix $\matr{H}_{\PGq}$, as a better understanding of this
fundamental cone yields a better understanding of linear programming (LP)
decoding~\cite{Feldman:Wainwright:Karger:05:1} of this code.  Moreover, the
connection made by Koetter and Vontobel~\cite{Koetter:Vontobel:03:1,
Vontobel:Koetter:04:2} between iterative decoding and LP decoding suggests
that results for LP decoding have immediate implications for iterative
decoding. We will use the same notations and definitions
of~\cite{Vontobel:Smarandache:Kiyavash:Teutsch:Vukobratovic:05:1} that we
briefly review here. We let $\R$, $\Rp$, and $\Rpp$ be the set of real number,
the set of non-negative real numbers, and the set of positive real numbers,
respectively.

\begin{Definition}[\cite{Koetter:Vontobel:03:1,
                         Feldman:Wainwright:Karger:05:1}]
  \label{def:fundamental:cone:1}
  
  Let $\code{C}$ be an arbitrary binary linear code that is described by a
  parity-check matrix $\matr{H}$ of size $m \times n$. We let $\set{J} \defeq
  \set{J}(\matr{H}) \defeq \{ 1, \ldots, m \}$ and $\set{I} \defeq
  \set{I}(\matr{H}) \defeq \{1,2,\ldots,n\}$ be the set of row and column
  indices of $\matr{H}$, respectively. For each $j \in \set{J}$, we let
  $\set{I}_j \defeq \set{I}_j(\matr{H}) \defeq \big\{ i \in \set{I} \ | \
  h_{ji} = 1 \big\}$ and for each $i \in \set{I}$ we let $\set{J}_i \defeq
  \set{J}_i(\matr{H}) \defeq \big\{ j \in \set{J} \ | \ h_{ji} = 1 \big\}$. We
  define the {\em fundamental cone} $\fch{K}{H}$ of $\matr{H}$ to be the set
  of vectors $\vomega \in \R^n$ that satisfy
  \begin{align}
    \forall j \in \set{J}, \
      \forall i \in \set{I}_j:
        \quad
    \sum_{i' \in \set{I}_j \setminus \{ i \}}
      \omega_{i'}
       \geq
        \omega_{i}
    \quad\quad
    \text{ and }
    \quad\quad
    \forall i \in \set{I}:
        \quad
    \omega_i
       \geq 0.
         \label{eq:fund:cone:ineq}
  \end{align}
  Vectors in the fundamental cone will be called \emph{pseudo-codewords}. Note
  that two pseudo-codewords that are equal up to a positive scaling constant
  will be considered to be equivalent. The edges of the fundamental cone will
  be called {\em minimal pseudo-codewords}. It can be shown that all minimal
  pseudo-codewords stem from valid configurations in covers of the base Tanner
  graph, and that minimal pseudo-codewords that are
  \emph{unnormalized}~\cite{Koetter:Li:Vontobel:Walker:04:1} are equal (modulo
  $2$) to some codewords of the code $\code{C}$. \edefinition
\end{Definition}

Note that the fundamental cone is a function of the parity-check matrix
representing a code. Because of the equivalence of parity-check matrix and
Tanner graph, the fundamental cone can also be seen as a function of the
Tanner graph representing a code. Therefore, in order to emphasize the
dependence of minimal pseudo-codewords on the representation of the code, we
will talk about the minimal pseudo-codewords of a Tanner graph.

Note also that the fundamental cone is \emph{independent} of the specific
memoryless binary-input channel through which we are transmitting; however,
the influence of a pseudo-codeword on the LP decoding behavior is measured by
a \emph{channel-dependent} pseudo-weight. For the binary-input additive white
Gaussian noise channel, the AWGNC-pseudo-weight turns out to be $\wps(\vomega)
\defeq \wpsAWGNC(\vomega) \defeq
\frac{\sonenorm{\vomega}^2}{\twonorm{\vomega}^2}$ if $\vomega \in \Rp^n
\setminus \{ \vect{0} \}$ and $\wps(\vomega) \defeq \wpsAWGNC(\vomega) \defeq
0$ if $\vomega = \vect{0}$~\cite{Wiberg:96,
Forney:Koetter:Kschischang:Reznik:01:1, Koetter:Vontobel:03:1}; the formula
for the binary symmetric channel (BSC) pseudo-weight $\wpsBSC(\vomega)$ can be
found in~\cite{Forney:Koetter:Kschischang:Reznik:01:1};\footnote{Because of
space reasons we omit the rather lengthy definition of $\wpsBSC(\vomega)$;
however, in Sec.~\ref{sec:effective:minimal:pseudo:codewords:1} we will
discuss some of the consequences of the $\wpsBSC(\vomega)$ definition.}
finally, for the binary erasure channel, the BEC-pseudo-weight is
$\wpsBEC(\vomega) \defeq
|\supp(\vomega)|$\,~\cite{Forney:Koetter:Kschischang:Reznik:01:1}.

Let $\wpsmin(\matr{H})$ be the minimum AWGNC pseudo-weight of a parity-check
matrix $\matr{H}$. One can show that $\wpsmin(\matr{H}_{\PGq}) \geq q + 2$
(e.g.~using Th.~1 in \cite{Vontobel:Koetter:04:1}) and because this lower
bound matches the minimum Hamming weight, we actually know that
$\wpsmin(\matr{H}_{\PGq}) = q + 2$. Similarly, one can show that
$\wpsBSCmin(\matr{H}_{\PGq}) = q + 2$, and that $\wpsBECmin(\matr{H}_{\PGq}) =
q + 2$.

\begin{figure}
  \begin{center}
    \epsfig{file=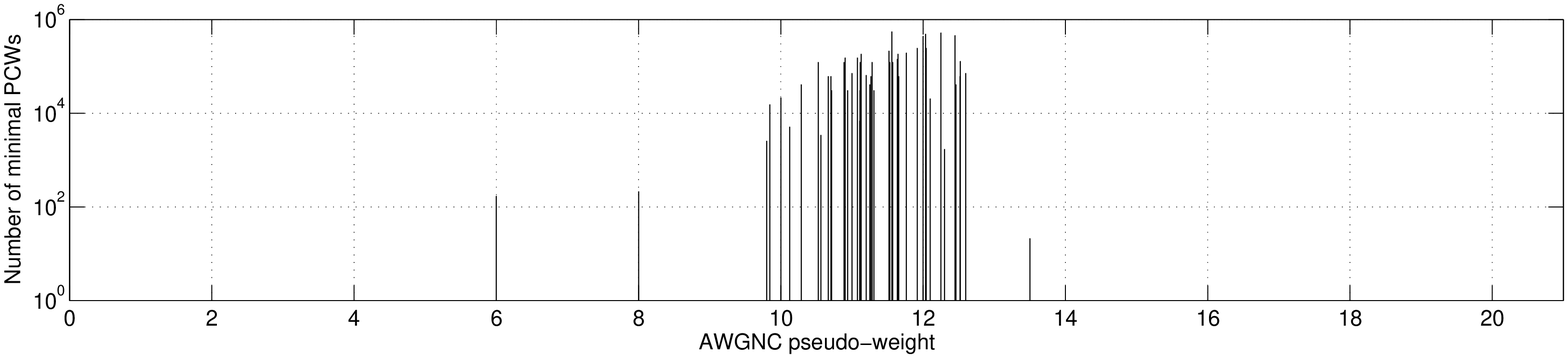,
            width=0.9\linewidth} \\[0.3cm]
    \epsfig{file=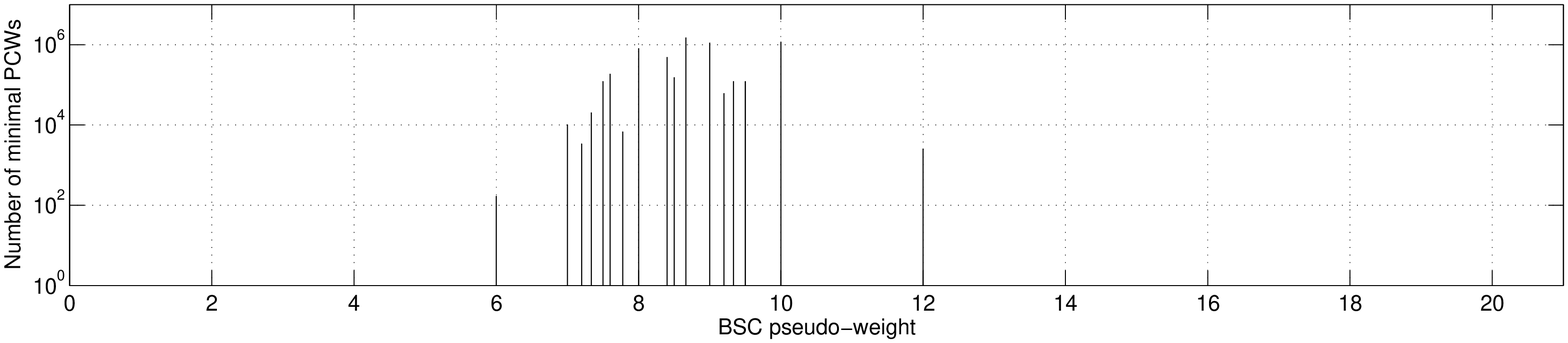,
            width=0.9\linewidth} \\[0.3cm]
    \epsfig{file=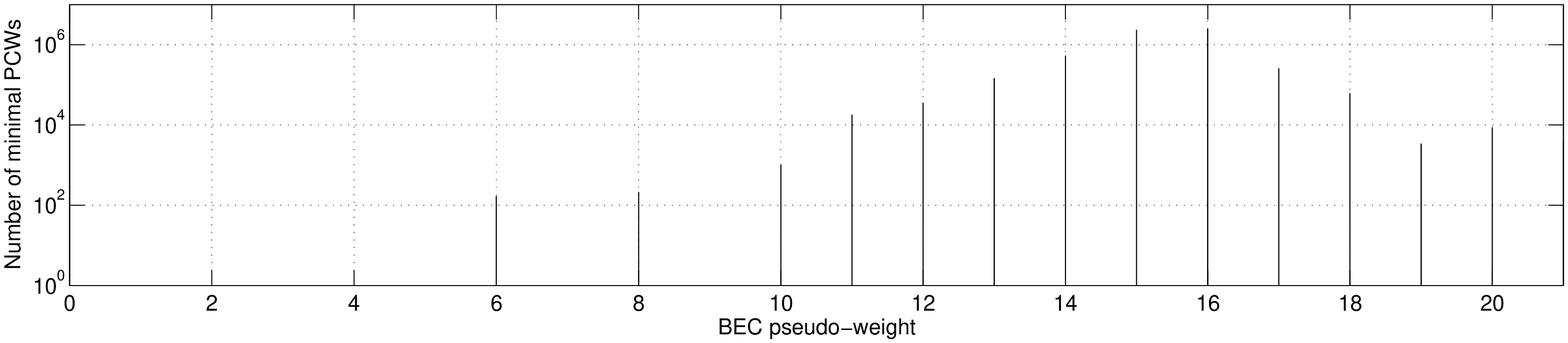,
            width=0.9\linewidth}
  \end{center}
  \caption{Histograms of the AWGNC, BSC, and BEC pseudo-weight of minimal
  pseudo-codewords of the $\PG{2}{4}$-based code, see
  also~\cite{Vontobel:Smarandache:Kiyavash:Teutsch:Vukobratovic:05:1}. (Note
  that the y-axis is logarithmic.)}
  \label{fig:pg:code:q:4:min:pseudo:codewords:histogram:1}
\end{figure}

\begin{Example}
  Consider the parity-check matrix $\matr{H}_{\PGq}$ for $q = 4$ and its
  associated Tanner
  graph. Fig.~\ref{fig:pg:code:q:4:min:pseudo:codewords:histogram:1} shows the
  histograms of the AWGNC, BSC, and BEC pseudo-weight of minimal
  pseudo-codewords of this Tanner graph.

  Without going into any details, it is apparent from
  Fig.~\ref{fig:pg:code:q:4:min:pseudo:codewords:histogram:1} that the
  influence of minimal pseudo-codewords can vary depending on the channel that
  is used. (For related observations about varying influences of minimal
  pseudo-codewords, see the discussion in~\cite{Haley:Grant:05:1}.)

  It is well-known that the support set of any pseudo-codeword is a stopping
  set~\cite{Di:Proietti:Telatar:Richardson:Urbanke:02:1} and that for any
  stopping set there exists a pseudo-codeword whose support set equals that
  stopping set. Therefore, the BEC pseudo-weight of a pseudo-codeword equals
  the size of a certain stopping set and so the work by Kashyap and
  Vardy~\cite{Kashyap:Vardy:03:1} on (minimal) stopping sets for
  finite-geometry-based codes has implications for our setup, in particular
  when studying the BEC pseudo-weight. \eexample
\end{Example}

\begin{Definition}
  \label{def:type:of:vector:1}
  Let $\vomega \in \Rp^n$. We call $\vtype \defeq \vtype(\vomega) =
  (t_{\ell}(\vomega))_{\ell \in \Rp}$ the type of $\vomega$, where
  $\type_{\ell} \defeq \type_{\ell}(\vomega)$ is the number of components of
  the vector $\vomega$ that are equal to $\ell$. (Note that in the following
  we do not assume that $\ell$ is a non-negative integer, only that it is a
  non-negative real number.) \edefinition
\end{Definition}

It follows from this definition that only finitely many $\type_{\ell}$'s are
non-zero and that $\sum_{\ell} \type_{\ell} = |\set{I}|=n$ for any $\vomega \in
\Rp^n$. Moreover, because $\onenorm{\vomega} = \sum_{\ell} \ell \type_{\ell}$,
$\twonorm{\vomega}^2 = \sum_{\ell} \ell^2 \type_{\ell}$, and $|\supp(\vomega)|
= \sum_{\ell > 0} \type_{\ell}$ we have
\begin{align*}
  \wps(\vomega)
    &= \frac{\left(\sum_{\ell} \ell \type_{\ell} \right)^2}
            {\sum_{\ell} \ell^2 \type_{\ell}}
  \quad\quad
  \text{ and }
  \quad\quad
  \wpsBEC(\vomega)
     = \sum_{\ell > 0} \type_{\ell}.
\end{align*}
If $\tilde \vomega = \alpha \cdot \vomega$ for some $\alpha \in \Rpp$ then its
type $\tilde \vtype \defeq \vtype(\tilde \vomega)$ is such that $\tilde
\type_{\alpha \ell} = \type_{\ell}$ for all $\ell$.

The rest of this paper is structured as follows.  Whereas in
Sec.~\ref{sec:bounds:awgnc:pseudo:weight:1} we will discuss bounds on the
AWGNC pseudo-weight, in Sec.~\ref{sec:effective:minimal:pseudo:codewords:1} we
will investigate the so-called effectiveness of minimal pseudo-codewords, and
in Sec.~\ref{sec:structure:of:minimal:pseudo-codewords:1} we will study the
structure of minimal pseudo-codewords. Finally, in
Sec.~\ref{sec:conclusions:1} we offer some conclusions.

\section{Bounds on the AWGNC Pseudo-Weight}
\label{sec:bounds:awgnc:pseudo:weight:1}

In this section we present some bounds on the AWGNC pseudo-weight, in
particular we present bounds that depend only on the type of a
pseudo-codeword.

\begin{Lemma}
  \label{lemma:awgnc:pseudo:weight:lower:bound:2:1}

  Let $\vect{\vomega} \in \Rp^n$ be a vector. If its type $\vtype =
  \vtype(\vomega)$ is such that only $\type_0$, $\type_1$, and $\type_2$ are
  non-zero, then
  \begin{align*}
    w_p(\vect{\vomega})
      &\geq \max \left\{\frac{15}{16} \type_1
            +
            \frac{12}{16} \type_2, ~~\frac{3}{4}\type_1
            +
            \type_2\right \}.
   \end{align*}
\end{Lemma}

\begin{Proof}
  Using the well-known bound $\sqrt{4\type_1 \type_2} \leq \frac{\type_1 + 4
  \type_2}{2}$, i.e.~$\frac{\type_1 \type_2}{\type_1 + 4 \type_2} \leq
  \frac{1}{16}(\type_1 + 4 \type_2)$, we obtain $w_p(\vect{\vomega}) =
  \frac{(\type_1 + 2 \type_2)^2}{\type_1 + 4 \type_2} = \type_1 + \type_2 -
  \frac{\type_1 \type_2}{\type_1 + 4 \type_2} \geq \type_1 + \type_2-
  \frac{\type_1 + 4\type_2}{16} = \frac{15\type_1}{16} +
  \frac{12\type_2}{16}$. For the second inequality we have
  $w_p(\vect{\vomega}) = \frac{(\type_1 + 2 \type_2)^2}{\type_1 + 4 \type_2}
  =\type_2+ \frac{\type_1(\type_1+ 3 \type_2)}{\type_1 + 4 \type_2}\geq
  \frac{3}{4}\type_1+\type_2.$
\end{Proof}

\begin{Lemma}
  \label{lemma:awgnc:pseudo:weight:lower:bound:1:1}

  Let $\vomega \in \Rp^n$ and let $\eta \neq 0$ be some arbitrary real
  number. Then
  \begin{align*}
    \wps(\vomega)
      &\geq
         \frac{2 \eta \onenorm{\vomega}
               -
               \twonorm{\vomega}^2}
              {\eta^2}
       = \frac{\sum\limits_{i=1}^{n} \omega_i (2 \eta - \omega_i)}
              {\eta^2}
  \end{align*}
  with equality if and only if $\vomega = \vect{0}$ or $\eta =
  \twonorm{\vomega}^2 / \onenorm{\vomega}$.
\end{Lemma}

\begin{Proof}
  If $\vomega = \vect{0}$ then the statement is certainly true, so let us
  assume that $\vomega \neq \vect{0}$. The square of any real number is
  non-negative, therefore
  \begin{align*}
   \left(
     \eta \onenorm{\vomega} - \twonorm{\vomega}^2
   \right)^2
   \geq 0,
  \end{align*}
  with equality if and only if $\eta = \twonorm{\vomega}^2 /
  \onenorm{\vomega}$. Multiplying out and rearranging we obtain
  \begin{align*}
    \eta^2 \onenorm{\vomega}^2
      & \geq
          2 \eta \onenorm{\vomega} \twonorm{\vomega}^2
          -
          \twonorm{\vomega}^4.
  \end{align*}
  Finally, dividing by $\eta^2 \twonorm{\vomega}^2$ and using the definition
  of $\wps(\vomega)$, we obtain the desired result.
\end{Proof}

\begin{Corollary}
  \label{cor:awgnc:pseudo:weight:lower:bound:1:2}

  Let $\vomega \in \Rp^n$, let $\vtype \defeq \vtype(\vomega)$ be the type of
  $\vomega$, and let $\eta \neq 0$ be some arbitrary real number. Then
  \begin{align*}
    \wps(\vomega)
      &\geq \sum_{\ell}
              \beta_{\ell} \type_{\ell}
    \quad\quad
    \text{ with }
    \quad\quad
    \beta_{\ell}
       = \frac{\ell (2\eta - \ell)}
              {\eta^2}
       = 1
         -
         \left(
           1
           -
           \frac{\ell}{\eta}
         \right)^2.
  \end{align*}
\end{Corollary}

\begin{Proof}
  The result follows immediately from
  Th.~\ref{lemma:awgnc:pseudo:weight:lower:bound:1:1}.
\end{Proof}

Note that choosing $\eta = 4/3$ in
Cor.~\ref{cor:awgnc:pseudo:weight:lower:bound:1:2} yields $\beta_0 = 0$,
$\beta_1 = 15/16$, and $\beta_2 = 12/16$, and that choosing $\eta = 2$ in
Cor.~\ref{cor:awgnc:pseudo:weight:lower:bound:1:2} yields $\beta_0 = 0$,
$\beta_1 = 3/4$, and $\beta_2 = 1$. This recovers
Lemma~\ref{lemma:awgnc:pseudo:weight:lower:bound:2:1}.

\begin{Corollary}
  \label{cor:awgnc:pseudo:weight:lower:bound:1:3}

  Let $\vomega \in \Rp^n$ and let $\vtype \defeq \vtype(\vomega)$. Moreover,
  let $r$ be the ratio of the largest positive $\ell$ such that $\type_{\ell}$
  is non-zero and the smallest positive $\ell$ such that $\type_{\ell}$ is
  non-zero. Then we have the lower bound
  \begin{align*}
    \wps(\vomega)
      &\geq
         \frac{4r}
              {(r+1)^2}
         \cdot
         |\supp(\vomega)|.
  \end{align*}
  (This bound was also obtained by Wauer~\cite{Wauer:05:1} using a different
  derivation.)
\end{Corollary}

\begin{Proof}
  Let $m$ be the largest positive $\ell$ such that $\type_{\ell}$ is non-zero
  and let $m'$ be the smallest positive $\ell$ such that $\type_{\ell}$ is
  non-zero. These definitions obviously yield $r = m/m'$. Consider
  Cor.~\ref{cor:awgnc:pseudo:weight:lower:bound:1:2} with $\eta =
  \frac{m+m'}{2}$. We obtain $\wps(\vomega) \geq \sum_{\ell} \beta_{\ell}
  \type_{\ell}$ $(*)$ with $\beta_{\ell} = 4 \ell \frac{m + m' - \ell}{(m +
  m')^2} = 1 - (1 - \frac{2 \ell}{m + m'})^2$. We observe that $\beta_{m'} =
  \beta_m = \frac{4 m m'}{(m + m')^2} = \frac{4 r}{(r+1)^2}$. Since
  $\beta_{\ell}$ is strictly concave in ${\ell}$ we must have $\beta_{\ell} >
  \beta_{m'} = \beta_{m} = \frac{4 r}{(r + 1)^2}$ for all $m' < \ell < m$.

  Choosing $\{ \beta'_{\ell} \}$ such that $\beta'_{\ell} \leq \beta_{\ell}$
  for all ${\ell}$, the above lower bound in~$(*)$ can be turned into the
  lower bound $\wps(\vomega) \geq \sum_{\ell} \beta'_{\ell} \type_{\ell}$
  because $\type_{\ell} \geq 0$ for all $\ell$. We choose $\beta'_{\ell}
  \defeq \frac{4 r}{(r+1)^2}$ for all $m' \leq \ell \leq m$ and $\beta'_{\ell}
  \defeq 0$ otherwise. The observations in the previous paragraph show that
  these are valid choices and we finish the proof by noting that
  \begin{align*}
    \wps(\vomega)
      &\geq
         \sum_{\ell}
           \beta'_{\ell} \type_{\ell}
       = \sum_{m' \leq \ell \leq m}
           \frac{4r}
              {(r+1)^2}
           \type_{\ell}
       = \frac{4r}
              {(r+1)^2}
           \sum_{m' \leq \ell \leq m}
             \type_{\ell}
       = \frac{4r}
              {(r+1)^2}
         \cdot
         |\supp(\vomega)|.
  \end{align*}
  \mbox{}\\[-0.75cm]
  \mbox{}
\end{Proof}

\begin{Theorem}
  \label{theorem:awgnc:pseudo:weight:lower:bound:2:1}

  Let $\matr{H} \defeq \matr{H}_{\PGq}$ and let $\vomega \in \fch{K}{H}$ be of
  type $\vtype$ with both $t_0$ non-negative, $t_1 \geq q+2$, $t_2$ positive,
  and $t_{\ell} = 0$ otherwise.\footnote{Let $\code{C}$ be the code defined by
  $\matr{H}$. If a pseudo-codeword is an unscaled
  pseudo-codeword~\cite{Koetter:Li:Vontobel:Walker:04:1} then it is equal
  (modulo $2$) to a codeword of $\code{C}$. Therefore, the number of odd
  components of an unscaled pseudo-codeword must either be zero or at least
  equal to the minimum Hamming weight of the code. So, if we actually know
  that $\vomega$ in the theorem statement is an unscaled pseudo-codeword then
  the requirement $\type_1 \geq q+2$ is equal to the requirement $\type_1 \geq
  1$. \label{footnote:unscaled:pseudo:codeword:1}} Then
  \begin{align*}
    \wps(\vomega)
      &\geq \frac{4}{3} (q+2).
  \end{align*}
\end{Theorem}

\begin{Proof}
  For any $i \in \set{I}$ we must have $\sum_{i' \in \set{I} \setminus \{ i
  \}} \omega_{i'} \overset{(*)}{=} \sum_{j \in \set{J}_i} \sum_{i' \in
  \set{I}_j \setminus \{ i \}} \omega_{i'} \overset{(**)}{\geq} \sum_{j \in
  \set{J}_i} \omega_i = (q+1) \omega_i$, where at step $(*)$ we used the fact
  that all variable nodes are at graph distance two from each other in the
  Tanner graph associated to $\matr{H}$, and where at step $(**)$ we used the
  inequalities in~\eqref{eq:fund:cone:ineq}. Adding $\omega_i$ to both sides
  we obtain $\sum_{i' \in \set{I}} \omega_{i'} \geq (q+2) \omega_i$. Now, fix
  an $i \in \set{I}$ for which $\omega_i = 2$ holds and express $\sum_{i' \in
  \set{I}} \omega_{i'}$ in terms of $\vtype$: it must hold that $\type_1 +
  2\type_2 \geq 2 (q+2)$, or, equivalently, $\type_2 \geq q + 2 -
  \type_1/2$. For any $\eta \neq 0$ we obtain
  \begin{align*}
    \wps(\vomega)
      &\overset{(*)}{\geq}
         \frac{(2\eta - 1) \type_1 + (4\eta - 4) \type_2}
              {\eta^2}
       \overset{(**)}{\geq}
         \frac{(2\eta - 1) \type_1 + (4\eta - 4) (q + 2 - \type_1/2)}
                 {\eta^2} \\
      &=    \frac{\type_1 + (4\eta - 4) (q + 2)}
                 {\eta^2},
  \end{align*}
  where at step $(*)$ we used
  Cor.~\ref{cor:awgnc:pseudo:weight:lower:bound:1:2} and at step $(**)$ we
  used the inequality on $\type_2$ that we just found above. Using the
  assumption that $\type_1 \geq q+2$ from the theorem statement we get
  $\wps(\vomega) \geq \frac{(4\eta - 3) (q + 2)} {\eta^2}$. The right-hand
  side of this expression is maximized by $\eta^{*} = \frac{3}{2}$: inserting
  this value yields the lower bound in the theorem statement.
\end{Proof}

A possible goal for future research is to weaken the assumptions about
$\type_1$ in the theorem statement without weakening the lower bound on the
AWGNC pseudo-weight of pseudo-codewords that are not (multiples of) codewords:
in light of Footnote~\ref{footnote:unscaled:pseudo:codeword:1} it would be
desirable to prove that the same lower bound holds also if $\vomega$ is an
unscaled pseudo-codeword with $\type_1 = 0$ and which is not a multiple of a
codeword.

Note that the above theorem can be generalized to the setup where $\vomega \in
\fch{K}{H}$ has type $\vtype$ with $t_0$ non-negative, $t_m$ positive for some
integer $m \geq 2$, $t_{\ell}$ non-negative for $1 \leq \ell \leq m-1$,
$t_{\ell} = 0$ for $\ell \geq m+1$, and $\sum_{\text{odd } \ell} t_{\ell} \geq
q+2$. Then $\wps(\vomega) \geq\frac{m^2}{m^2-m+1} (q+2)$.

\begin{figure}
  \begin{center}
    \epsfig{file=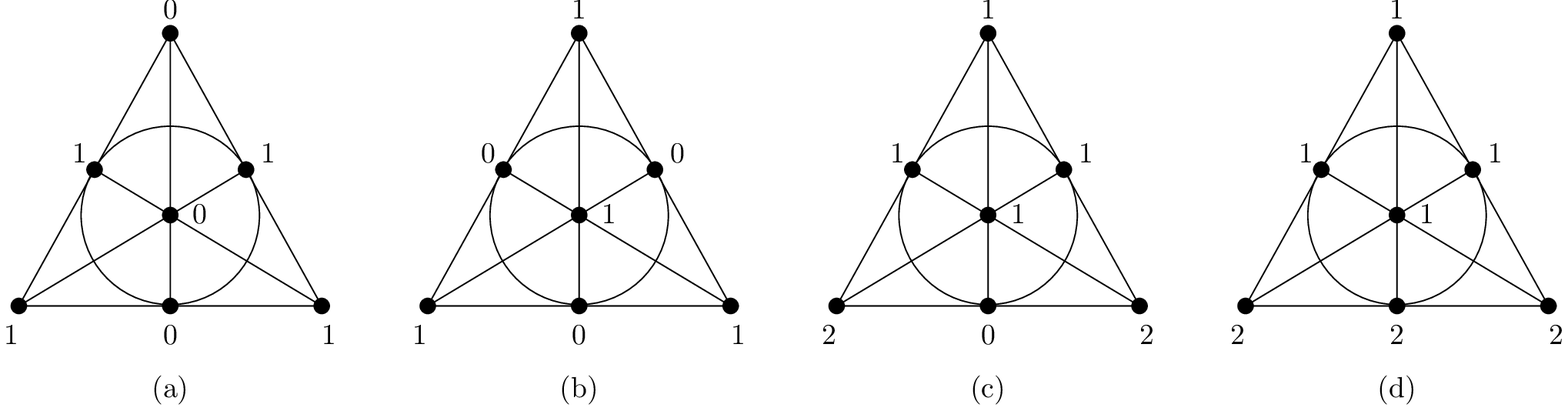, width=0.6\linewidth}
    \quad\quad\quad\quad
    \epsfig{file=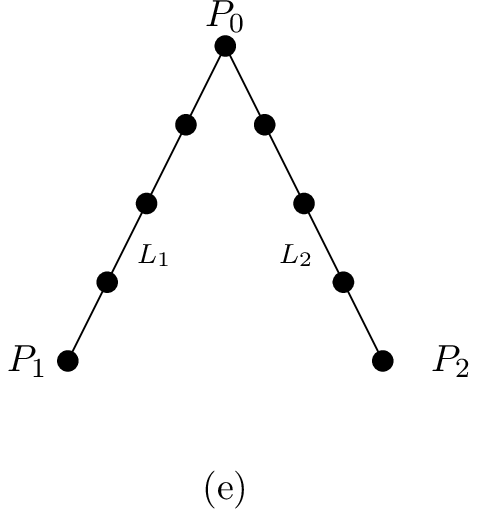, width=0.14\linewidth}
  \end{center}
  \caption{(a)-(d): Codewords and pseudo-codewords used in
           Ex.~\ref{ex:adding:codewords:and:switching:1}. (e): Part of
           $\PG{2}{4}$ discussed in Ex.~\ref{ex:structure:of:minimal:pcw:1}}
  \label{fig:pg:2:2:pcw:1:pg:2:4:pcw:1}
\end{figure}

\begin{Example}
  \label{ex:adding:codewords:and:switching:1}

  One can exhibit minimal pseudo-codewords whose AWGNC pseudo-weight matches
  the leading-term behavior of the lower bound in
  Th.~\ref{theorem:awgnc:pseudo:weight:lower:bound:2:1} (when $q$
  grows). Consider first the case $q = 2$. The projective plane for $q = 2$ is
  shown in Fig.~\ref{fig:pg:2:2:pcw:1:pg:2:4:pcw:1}~(a): it has $7$ points and
  $7$ lines and we consider the points to be variables and the lines to be
  checks. Fig.~\ref{fig:pg:2:2:pcw:1:pg:2:4:pcw:1}~(a and b) shows two
  codewords of weight $q + 2 = 4$; note that their supports overlap in
  $\frac{q + 2}{2} = 2$ positions. Adding these two codewords together yields
  the pseudo-codeword shown in
  Fig.~\ref{fig:pg:2:2:pcw:1:pg:2:4:pcw:1}~(c). Switching the zero value into
  a two results in the pseudo-codeword in
  Fig.~\ref{fig:pg:2:2:pcw:1:pg:2:4:pcw:1}~(d); it can be checked that this
  pseudo-codeword is actually a minimal pseudo-codeword. It has AWGNC
  pseudo-weight $6.25$, whereas the lower bound in
  Th.~\ref{theorem:awgnc:pseudo:weight:lower:bound:2:1} is $5.33$.

  Similarly, in the case of $q = 4$ it is possible to start with two codewords
  of weight $q+2 = 6$ whose supports overlap in $\frac{q + 2}{2} = 3$
  positions. After adding them and switching two zeros (that are
  specifically chosen and lie on the same line) into two twos, one gets a
  minimal pseudo-codeword of AWGNC pseudo-weight $9.85$, whereas the lower
  bound in Th.~\ref{theorem:awgnc:pseudo:weight:lower:bound:2:1} is $8.00$.
 
  In general, we conjecture that for any $q = 2^s$, where $s$ is a positive
  integer, it is possible to construct a minimal pseudo-codeword of type
  $\vtype$ with $\type_1 = q + 2$ and $\type_2 = \frac{q}{2} + s + 1$ and
  $t_{\ell} = 0$ for $\ell \notin \{ 0, 1, 2 \}$: take two codewords of weight
  $q+2$ whose supports overlap in $\frac{q+2}{2} = \frac{q}{2} + 1$ positions
  and switch $s$ zeros (that are specifically chosen) into $s$ twos. The
  points corresponding to these $s$ twos (together with the lines through
  them) should then form a simplex.  These pseudo-codewords have weight
  \begin{align*}
    \wps(\vomega)
      &= \frac{\sonenorm{\vomega}^2}{\twonorm{\vomega}^2}
       = \frac{4}{3}
         \cdot
         (q+2)
         \cdot
         \frac{1 + f(q)}
              {1 + \frac{f(q)}{3(1+f(q))}},
  \end{align*}
  where $f(q) = \frac{\log_2(q)}{q+2}$. (We wrote the last term on the
  right-hand side such that it is readily apparent that it is not smaller than
  $1$, i.e.~the bound in Th.~\ref{theorem:awgnc:pseudo:weight:lower:bound:2:1}
  is clearly satisfied.) \eexample
\end{Example}

\section{Effective Minimal Pseudo-Codewords}
\label{sec:effective:minimal:pseudo:codewords:1}

The BSC can be seen as a binary-input AWGNC where the values at the output are
quantized to $+1$ or $-1$. It follows that the components of the
log-likelihood vector $\vlambda$ can only take on two values, namely $+L$ and
$-L$, where $L$ is a positive constant that depends on the bit flipping
probability of the BSC. Because of this quantization, there are certain
effects that happen for the BSC that cannot happen for the AWGNC. Before
continuing, it is worthwhile to recall what the meaning of the BSC
pseudo-weight $\wpsBSC(\vomega)$ of a pseudo-codeword
$\vomega$~\cite{Forney:Koetter:Kschischang:Reznik:01:1} is: $\lceil
\wpsBSC(\vomega) / 2 \rceil$ is the minimum number of bit flips required (upon
sending the zero codeword) to make a decoding error to $\vomega$; moreover,
these bit flips must happen at appropriate positions.

\begin{Definition}
  Fix a memoryless binary-input channel and let $\set{L}^{(n)} \subseteq (\R
  \cup \{ \pm \infty \})^n$ be the set of all possible log-likelihood ratio
  vectors.\footnote{For the AWGNC we have $\set{L}^{(n)} = \R^n$, for the BSC
  we have $\set{L}^{(n)} = \{ \pm L \}^n$ for some $L \geq 0$, and for the BEC
  we have $\set{L}^{(n)} = \{ -\infty, 0, +\infty \}^n$.} Moreover, let us fix
  a parity-check matrix $\matr{H}$ and let $\Mps(\fch{K}{H})$ be the set of
  minimal pseudo-codewords. A minimal pseudo-codeword $\vomega \in
  \Mps(\fch{K}{H})$ is called \emph{effective of the first kind} for that
  particular channel if there exists a $\vlambda \in \set{L}^{(n)}$ such that
  $\langle \vomega, \vlambda \rangle < 0$ and $\langle \vomega', \vlambda
  \rangle \geq 0$ for all $\vomega' \in \Mps(\fch{K}{H}) \setminus \{ \vomega
  \}$. A minimal pseudo-codeword $\vomega \in \Mps(\fch{K}{H})$ is called
  \emph{effective of the second kind} for that particular channel if there
  exists a $\vlambda \in \set{L}^{(n)}$ such that $\langle \vomega, \vlambda
  \rangle \leq 0$ and $\langle \vomega', \vlambda \rangle \geq 0$ for all
  $\vomega' \in \Mps(\fch{K}{H}) \setminus \{ \vomega \}$. (Obviously, a
  minimal pseudo-codeword that is effective of the first kind is also
  effective of the second kind.)  \edefinition
\end{Definition}

Let $\set{L}^{(n)}_{\vect{0}} \subseteq \set{L}^{(n)}$ be the set where LP
decoding decides in favor of the codeword $\vect{0}$. From the above
definition it follows that a minimal pseudo-codeword shapes the set
$\set{L}^{(n)}_{\vect{0}}$ if and only if it is an effective minimal
pseudo-codeword. More precisely, in the case where a minimal pseudo-codeword
$\vomega$ is effective of the first kind then there exists at least one
$\vlambda \in \set{L}^{(n)}$ where $\vomega$ wins against all other minimal
pseudo-codewords (and the zero codeword). However, in the case where $\vomega$
is effective of the second kind we are guaranteed that there is at least one
$\vlambda \in \set{L}^{(n)}$ were $\vomega$ is involved in a tie; if and how
often $\vomega$ wins against all other minimal pseudo-codewords (and the zero
codeword) depends on how ties are resolved.

\begin{Theorem}
  For the binary-input AWGNC and any parity-check matrix $\matr{H}$ all
  minimal pseudo-codewords of $\fch{K}{H}$ are effective of the first kind.
\end{Theorem}

\begin{Proof}
  This follows from some simple geometric considerations.
\end{Proof}

\mbox{}

As the following observations show, for channels other than the AWGNC not all
minimal pseudo-codeword need to be effective of the first or second kind.

\begin{Theorem}
  \label{theorem:bsc:correction:capabilities:1}

  Consider data transmission over a BSC using the code defined by $\matr{H}
  \defeq \matr{H}_{\PGq}$. LP decoding can correct any pattern of
  $\frac{q}{2}$ bit flips and no pattern of more than $q$ bit flips.
\end{Theorem}

\begin{Proof}
  It can be shown that the BSC pseudo-weight of any pseudo-codeword in
  $\fch{K}{H}$ is at least $q + 2$. Therefore LP decoding can correct at least
  $\lfloor \frac{q + 2 - 1}{2} \rfloor = \frac{q}{2}$ bit flips.

  Let us now show that LP decoding can correct at most $q$ bit flips. Remember
  that a necessary condition for LP decoding to decode a received
  log-likelihood vector $\vlambda$ to the zero codeword is that $\langle
  \vomega, \vlambda \rangle \geq 0$ for all $\vomega \in
  \fch{K}{H}$.\footnote{Note that this is usually not a sufficient condition
  for correct decoding, e.g.~in the case where ties are resolved randomly.}
  Assume that we are transmitting the zero codeword and that $e$ bit flips
  happened. Hence $e$ components of $\vlambda$ are equal to $-L$ and $n-e$
  components of $\vlambda$ are equal $+L$. It can easily be checked that the
  following $\vomega$ is in $\fch{K}{H}$: let $\omega_i \defeq 1$ if
  $\lambda_i = -L$ and $\omega_i \defeq 1/q$ otherwise.\footnote{This can be
  seen as a generalization of the so-called canonical
  completion~\cite{Koetter:Vontobel:03:1}, however instead of assigning values
  according to the graph distance with respect to a single node, we assign
  values according to the graph distance with respect to the set of nodes
  where $\lambda_i$ is negative. Note that special property of the Tanner
  graph of $\matr{H}$: all variable nodes are at graph distance $2$ from each
  other.} For this $\vomega$, the condition $\langle \vomega, \vlambda \rangle
  \geq 0$ translates into $e \cdot (-L) + (n-e) \cdot (1/q) \cdot (+L) \geq
  0$, i.e.~$e \leq \frac{n}{q+1} = \frac{q^2 + q + 1}{q+1} = q +
  \frac{1}{q+1}$. Rounding down we obtain $\lfloor e \rfloor = \lfloor q +
  \frac{1}{q+1} \rfloor = q$.
\end{Proof}

\begin{Corollary}
  \label{cor:bounds:on:wpsBSC:for:effective:pcw:1}

  Consider the code defined by $\matr{H} \defeq \matr{H}_{\PGq}$. For the BSC,
  a necessary condition for a minimal pseudo-codeword $\vomega$ of
  $\fch{K}{H}$ to be effective of the second kind is that $q + 2 \leq
  \wpsBSC(\vomega) \leq 2q + 2$.
\end{Corollary}

For $q = 4$ it turns out that $\fc{K}(\matr{H}_{\PG{2}{4}})$ has minimal
pseudo-codewords with BSC pseudo-weight equal to $12$. (These minimal
pseudo-codewords have type $\vtype$ with $\type_2 = 1$, $\type_1 = 12$,
$\type_0 = 8$, and $t_{\ell} = 0$ otherwise.)
Cor.~\ref{cor:bounds:on:wpsBSC:for:effective:pcw:1} clearly shows that these
cannot be effective of the second kind for the BSC, since, for $q = 4$, any
effective minimal pseudo-codeword of the second kind must fulfill $6 \leq
\wpsBSC(\vomega) \leq 10$.

Judging from Fig.~\ref{fig:pg:code:q:4:min:pseudo:codewords:histogram:1} it
also seems --- as far as AWGNC and BSC pseudo-weight are comparable --- that
soft information is quite helpful for the LP decoder when decoding the code
$\code{C}_{\PG{2}{4}}$ defined by $\matr{H}_{\PG{2}{4}}$.

One can also make interesting statements about the effectiveness of minimal
pseudo-codewords for the BEC; however, we postpone this discussion to a longer
version of the present paper.

\section{The Structure of Minimal Pseudo-Codewords}
\label{sec:structure:of:minimal:pseudo-codewords:1}

In this section we discuss the geometry of minimal pseudo-codewords. The
minimum weight of $\code{C}_{\PGq}$, $q$ a prime power, is $q + 2$ and
codewords that achieve this minimum weight correspond to point-line
configurations in the projective plane that have been studied by several
authors. Let us introduce some notation and results from finite geometries,
cf.~e.g.~\cite{Hirschfeld:79:1}. A $k$-arc in $\PGq$ is a set of $k$ points no
three of which are collinear. A $k$-arc is complete if it is not contained in
a $(k+1)$-arc. The maximum number of points that a $k$-arc can have is denoted
by $m(2,q)$, and a $k$-arc with this number of points is called an oval (in
the case where $q$ is even this is sometimes also called a hyper-oval). One
can show that $m(2,q) = q + 2$ for $q$ even and $m(2,q) = q + 1$ for $q$
odd. One can make the following two interesting observations for the case $q$
even. Firstly, if two ovals have more than half their points in common, then
these two ovals coincide. Secondly, if a $q$-arc is contained in an oval then
the number of such ovals is one if $q > 2$ and two if $q = 2$.

It turns out that in the case $q$ even, the codewords with minimal weight are
$q+2$-arcs and therefore ovals. However, whereas the classification of ovals
for odd $q$ is simple (they all correspond to conics), the ovals for even $q$
are not classified that easily. For even $q$, one says that an oval is regular
if it comprises the points of a conic and its nucleus; one can show that for
$q = 2^s$, irregular ovals exist for $s = 5$ and $s \geq 7$. It turns out that
the classification for irregular ovals is highly non-trivial. So, given that
even the classification of the codewords of minimal weight is difficult, it is
probably hopeless to obtain a complete classification of the minimal codewords
and minimal pseudo-codewords of codes defined by $\matr{H}_{\PGq}$, however it
is an interesting goal to try to understand as much as possible about the
structure of these codewords and pseudo-codewords.

From now on, $q$ will always be even, i.e.~a power of two. Before we state our
conjecture about the structure of minimal pseudo-codewords, let us first look
at an example.

\begin{Example}
  \label{ex:structure:of:minimal:pcw:1}
  
  Let $q=4$. Then we can find a minimal pseudo-codeword $\vomega$ whose type
  $\vtype$ is $\type_0 = 8$, $\type_1 = 8$, $\type_2 = 5$, and $\type_{\ell} =
  0$ otherwise. This pseudo-codeword can be obtained using a procedure similar
  to the one used in Ex.~\ref{ex:adding:codewords:and:switching:1}. Firstly,
  on has to add two vectors $\vx^{(1)}$ and $\vx^{(2)}$ of weight $6$ whose
  supports overlap in two positions. This yields a pseudo-codeword $\tilde
  \vomega$ of type $\tilde \vtype$ with $\tilde \type_0 = 11$, $\tilde \type_1
  = 8$, $\tilde \type_2 = 2$, and $\tilde \type_{\ell} = 0$ otherwise.
  Secondly, one has to switch three zeros (that were appropriately chosen)
  into three twos.

  Let us analyze this procedure. Since a minimal pseudo-codeword corresponds
  to an edge of the fundamental cone, it is clear that the inequalities
  in~\eqref{eq:fund:cone:ineq} that are fulfilled with equality must form a
  system of linear equations whose rank is $21 - 1 = 20$. We start with two
  minimal codewords $\vx^{(1)}$ and $\vx^{(2)}$ that each yield a system of
  linear equations whose rank is $21 - 1 = 20$. These two codewords have been
  chosen such that their sum $\tilde \vomega$ yields a system of linear
  equations whose rank is $21 - 2 = 19$.

  To find the three zeros that we have to switch, we proceed as follows. It
  turns out that in the projective plane $\PG{2}{4}$ there are two lines, say
  $L_1$ and $L_2$, such that all the entries of $\tilde \vomega$ that
  correspond to the points on these two lines are zero. Let $P_0$ be the
  intersection point of these two lines,
  cf.~Fig.~\ref{fig:pg:2:2:pcw:1:pg:2:4:pcw:1}~(e). There exists a point $P_1$
  on $L_1$ and a point $P_2$ on $L_2$ such that modifying $\tilde \vomega$ by
  assigning them the same value $\alpha \geq 0$ yields a vector in the
  fundamental cone, as long as $\alpha$ is not too large. In fact, for $\alpha
  > 2$ the vector is outside the fundamental cone, and for $\alpha = 2$ it
  yields a vector that is a pseudo-codeword and that yields a system of
  equations of rank $21 - 1 = 20$, i.e.~it is a minimal
  pseudo-codeword. \eexample
\end{Example}

\begin{Conjecture}
  For the Tanner graph defined by $\matr{H}_{\PGq}$ every minimal
  pseudo-codeword is a sum of a few minimal pseudo-codewords with a change of
  one or two low-value components such that they become the large components
  in the equations associated to the lines that pass through them.

  Hence, to find minimal pseudo-codewords, we have to take sums of two minimal
  pseudo-codewords that give rank $n-2$ (if possible, lower otherwise) and
  change one component that is not significant into a significant one. We call
  a component significant if it is the sum of the other components that belong
  to a line passing through the point, for most of such lines.
\end{Conjecture}

Answering positively the following conjecture would result in a much better
understanding of the minimal pseudo-codewords in general and of the so-called
AWGNC pseudo-weight spectrum
gap~\cite{Vontobel:Smarandache:Kiyavash:Teutsch:Vukobratovic:05:1}, in
particular.

\begin{Conjecture}
  Let $\matr{H} \defeq \matr{H}_{\PGq}$ and consider the pseudo-codewords that
  have minimal AWGNC pseudo-weight among all minimal pseudo-codewords that are
  not multiples of minimal codewords. We conjecture that the type $\vtype$ of
  these pseudo-codewords is such that $\type_0$ is non-negative, $\type_1$ is
  positive, $\type_2$ is positive, and $\type_{\ell} = 0$ otherwise. (If this
  conjecture is not true, find the the smallest $\tilde \ell$ such that these
  pseudo-codewords have type $\vtype$ with $t_{\ell} \geq 0$ for $\ell \in \{
  0, 1, \ldots, \tilde \ell \}$ and $\type_{\ell} = 0$ otherwise.)
\end{Conjecture}

\section{Conclusions}
\label{sec:conclusions:1}

In this paper we have gathered some new facts about minimal pseudo-codewords
of codes derived from finite projective planes. We have obtained a clearer
picture about the structure of these minimal pseudo-codewords, nevertheless
more work is required to get a sufficiently tight characterization of
them. Interestingly, in Sec.~\ref{sec:effective:minimal:pseudo:codewords:1},
we were able to use the canonical completion, a tool that so far has been very
useful for characterizing families of $(j,k)$-regular LDPC codes, with $j,~ k$
bounded when the block length goes to infinity, i.e.~code families where the
Tanner graph diameter grows with the block length.

In addition, because the AWGNC pseudo-weight spectrum gap seems to be large
for the codes considered in this paper, reflecting the fact that LP decoding
performs closely to ML decoding, LP decoding might be an interesting starting
point for obtaining a complete decoder for these codes, i.e.~a decoder that
finds the optimal codeword (or near-optimal codeword) with high probability
when $\vlambda$ is drawn according to the Gaussian distribution
$\mathcal{N}(0,\sigma^2)$, for some $\sigma^2$.

\bibliographystyle{ieeetr}

{\small 
\bibliography{/home/vontobel/references/references}
}

\end{document}